\documentclass{emulateapj}
\RequirePackage{epstopdf}

\slugcomment{Draft of \today}

\shorttitle{Holmberg II X-1 Paper}
\shortauthors{Lau et al.}

\usepackage{amsmath}
\usepackage{graphicx}
\usepackage[section] {placeins}
\usepackage{hyperref}
\usepackage[normalem]{ulem}
\usepackage{color}

\newcommand{\beq}{\begin{equation}}
\newcommand{\eeq}{\end{equation}}

\begin{document}

\title{First Detection of Mid-Infrared Variability from an Ultraluminous X-Ray Source Holmberg II X-1}

\author{R. M. Lau\altaffilmark{1,2},
M. Heida\altaffilmark{2},
M. M. Kasliwal\altaffilmark{2},
D. J. Walton\altaffilmark{3}}

\altaffiltext{1}{Jet Propulsion Laboratory, California Institute of Technology, 4800 Oak Grove Drive, Pasadena, CA 91109, USA}
\altaffiltext{2}{Division of Physics, Mathematics and Astronomy, Department of Astronomy, California Institute of Technology, Pasadena, CA 91125, USA}
\altaffiltext{3}{Institute of Astronomy, Madingley Road, CB3 0HA Cambridge, United Kingdom}

\begin{abstract}

We present mid-infrared (IR) light curves of the Ultraluminous X-ray Source (ULX) Holmberg II X-1 from observations taken between 2014 January 13 and 2017 January 5 with the \textit{Spitzer Space Telescope} at 3.6 and 4.5 $\mu$m in the \textit{Spitzer} Infrared Intensive Transients Survey (SPIRITS).  The mid-IR light curves, which reveal the first detection of mid-IR variability from a ULX, is determined to arise primarily from dust emission rather than from a jet or an accretion disk outflow. We derived the evolution of the dust temperature ($T_\mathrm{d}\sim600 - 800$ K), IR luminosity ($L_\mathrm{IR}\sim3\times10^4$ $\mathrm{L}_\odot$), mass ($M_\mathrm{d}\sim1-3\times10^{-6}$ $\mathrm{M}_\odot$), and equilibrium temperature radius ($R_\mathrm{eq}\sim10-20$ AU). A comparison of X-1 with a sample spectroscopically identified massive stars in the Large Magellanic Cloud on a mid-IR color-magnitude diagram suggests that the mass donor in X-1 is a supergiant (sg) B[e]-star. The sgB[e]-interpretation is consistent with the derived dust properties and the presence of the [Fe II] ($\lambda=1.644$ $\mu$m) emission line revealed from previous near-IR studies of X-1. We attribute the mid-IR variability of X-1 to increased heating of dust located in a circumbinary torus. It is unclear what physical processes are responsible for the increased dust heating; however, it does not appear to be associated with the X-ray flux from the ULX given the constant X-ray luminosities provided by serendipitous, near-contemporaneous X-ray observations around the first mid-IR variability event in 2014. Our results highlight the importance of mid-IR observations of luminous X-ray sources traditionally studied at X-ray and radio wavelengths.








\end{abstract}

\maketitle

\section{Introduction}

Ultraluminous X-ray sources (ULXs) are characterized by extreme X-ray luminosities ($L_X>10^{39}$ ergs $\mathrm{s}^{-1}$) and are primarily located outside the nuclear regions of their host galaxies (e.g. Fabbiano 2006). The energy output rates from ULXs exceeds the Eddington limit for accretion onto a $10$ M$_\odot$ black hole (Feng and Soria 2011), which requires super-Eddington accretion or intermediate mass black hole accretors ($M_\mathrm{IMBH}>100$ M$_\odot$; Colbert \& Mushotzky 1999). Intriguingly, recent observational studies have revealed a population of ULXs that exhibit X-ray pulsations consistent with rapidly rotating and magnetized neutron stars (Bachetti et al. 2014; Israel et al. 2016; F{\"u}rst et al. 2016). Accretion onto the neutron star in these ULXs must therefore occur at super-Eddington rates: ULX-1 in NGC 5907 shows peak luminosities of $\sim500$ times the neutron star Eddington limit (Isreal et al. 2016). Such extreme accretion rates are difficult to explain with current theoretical accretion models. Obtaining a better understanding of the mass transfer from the donor star onto the compact object companion is therefore crucial for determining the processes that drive the ultraluminous X-ray emission. 


Mid-infrared (mid-IR) wavelengths ($\lambda\sim3 - 30$ $\mu$m) can provide valuable insight into the circumstellar environment of X-ray binary systems by probing emission from warm ($\gtrsim100$ K) dust, accretion disk winds, and/or jets (e.g. Muno \& Mauerhan 2006; Moon et al. 2007; Rahoui et al. 2010; Mauerhan et al. 2010). Studies of ULXs at optical wavelengths are complicated by extinction and multiple emission components such as the accretion disk, the photosphere of the companion star, and/or surrounding nebulosity.

 Only a few observational studies exist on the mid-IR emission from ULXs (Berghea et al. 2010a, b; Dudik et al. 2016). Berghea et al. (2010a, b) present mid-IR photometry of the well-studied Holmberg II X-1 (hereafter referred to as X-1; Pakull \& Mirioni 2002; Lehmann et al. 2005; Tao et al. 2012; Walton et al. 2015) and analyze its full radio to X-ray spectral energy distribution (SED). Based on the radio and mid-IR fluxes of X-1, Berghea et al. (2010a, b) interpret the emission as arising from a jet; however, their radio and mid-IR data were taken over 10 years apart, which presents difficulties in taking variability into account in their modeling. Indeed, Cseh et al. (2014, 2015) report that radio emission from X-1 originating from a jet ejection event has faded by a factor $\sim7$ over the $\sim1.5$ yr period between 2012 Nov and 2014 May. Contemporaneous mid-IR/radio observations, long-term mid-IR monitoring, and consideration of alternative mid-IR emission mechanisms are therefore important for understanding the nature of the mid-IR counterpart of X-1. 

\begin{figure*}[t!]
	\centerline{\includegraphics[scale=1]{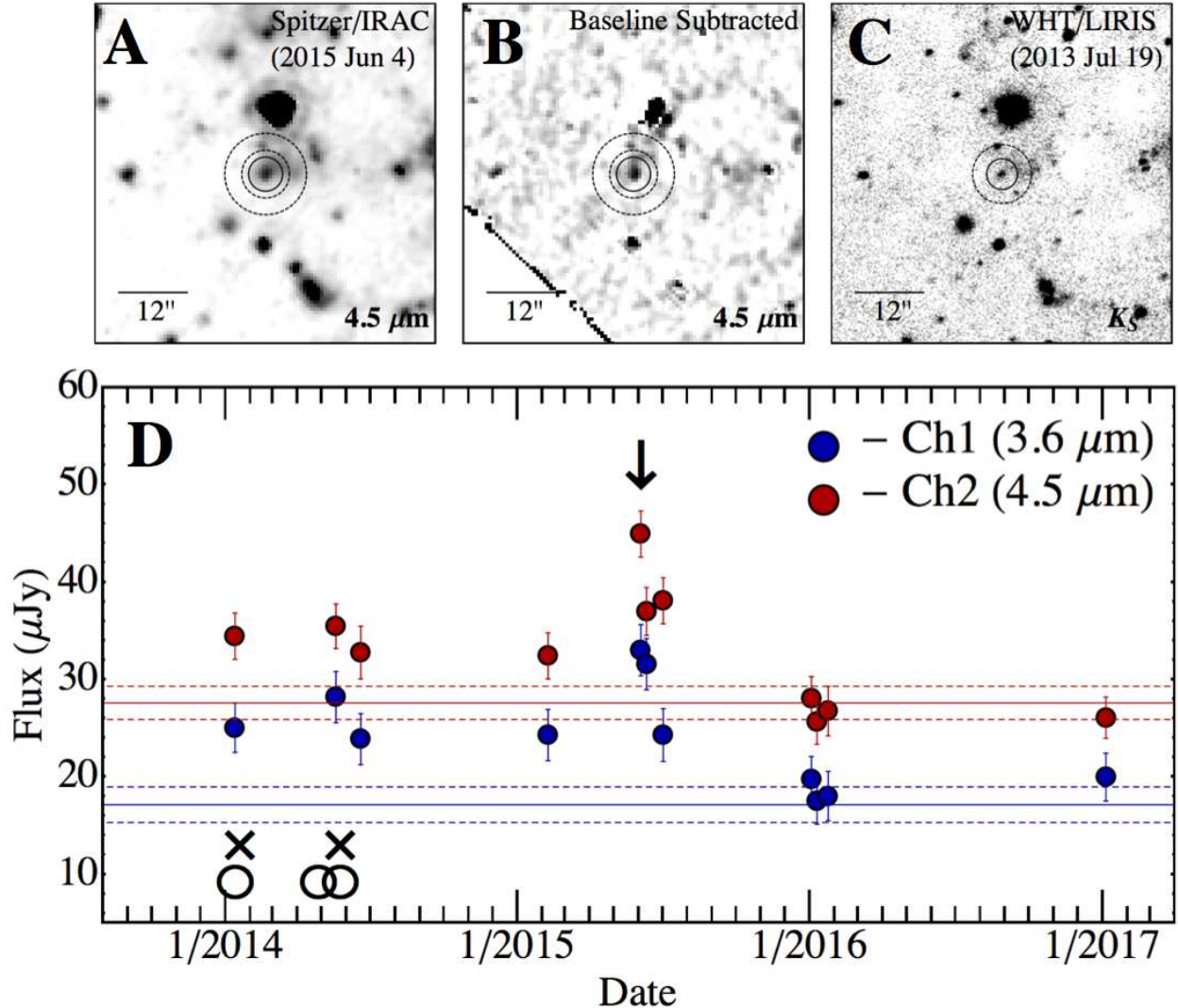}}
	\caption{(A) \textit{Spitzer}/IRAC 4.5 $\mu$m image centered on X-1 taken during peak measured flux on 2015 Jun 4 overlaid with the photometric aperture. (B) Baseline-subtracted image of (A) showing the flux variability from X-1. The baseline image was taken 2004 Oct 8. (C) WHT/LIRIS near-IR K$_S$-band image from Heida et al. (2014) with the same field of view as (A) overlaid with the photometric aperture used by Heida et al. (2014). (D) Mid-IR light curve of X-1 at 3.6 (blue) and 4.5 (red) $\mu$m. The solid lines indicate the mean baseline flux from observations taken on 2004 Oct 8 and 10, and the arrow corresponds to the date where the peak mid-IR flux was measured from X-1. The dashed lines and error-bars correspond to the $1\sigma$ flux uncertainty. Crosses and circles correspond to dates of X-ray and radio observations, respectively, by Cseh et al. (2015).}
	\label{fig:LC}
\end{figure*}

In this letter, we present 3.6 and 4.5 $\mu$m light curves of X-1 taken by the Infrared Array Camera (IRAC; Fazio et al. 2004) on board the warm \textit{Spitzer Space Telescope} (Werner et al. 2004; Gehrz et al. 2007). The multi-year baseline spanned by the mid-IR light curve allows us to perform a time-domain study of X-1 and investigate the origin of the mid-IR emission, its circumstellar environment, and the nature of the donor star.


\section{Mid-IR Imaging Observations from SPIRITS}
\label{Sec:Obs}

The SPitzer InfraRed Intensive Transient Survey (SPIRITS, PIDs 10136, 11063, \& 13053; Kasliwal et al. 2017) targets 194 nearby galaxies within 20 Mpc to a depth of 20 mag on the Vega system at 3.6 and 4.5 $\mu$m. Imaging observations with seven dithered 100 s exposures were performed in the Channel 1 and 2 bands of Spitzer/IRAC. X-1 was serendipitously included in observations of Holmberg II, one of the SPIRITS galaxies, and was observed at 10 epochs between Jan 2014 and Jan 2017 as indicated in Tab.~\ref{tab:Flux}. Archival mid-IR IRAC imaging data of Holmberg II taken 2004 Oct 8 and 2004 Oct 10 (PID 159) were utilized in this study. 


The mid-IR counterpart to X-1 is detected in both archival and SPIRITS images located at a position consistent with the X-ray and near-IR Epoch 2000 coordinates: RA - 08:19:28.99, Dec - 70:42:19.4 (Error - 0.7''; Swartz et al. 2004; Heida et al. 2014). The astrometric error for the \textit{Spitzer}/IRAC observations is $\sim0.6''$. \textit{Spitzer}/IRAC 4.5 $\mu$m imaging data of X-1 (Fig.~\ref{fig:LC}A), the baseline-subtracted image (Fig.~\ref{fig:LC}B), and a higher spatial resolution $K_S$-band image from Heida et al. (2014) taken from the William Herschel Telescope with the Long-slit Intermediate Resolution Infrared Spectrograph (WHT/LIRIS, Fig.~\ref{fig:LC}C) show that the mid- and near-IR counterparts are spatially coincident. A description of producing baseline-subtracted/difference images from SPIRITS and the appearance of false positives (e.g. subtraction of point spread functions at different field rotation angles) is provided in Kasliwal et al. (2017).


Aperture photometry was performed on the Post Basic Calibrated Data, where the flux was summed in a 5 pixel ($3''$) radius aperture centered on the mid-IR counterpart of X-1. Sky and the local nebular background was measured within an annulus from 7 to 12 pixels surrounding the source and subtracted from the total flux. The black circles in Fig.~\ref{fig:LC}A and B corresponds to the size and position of the aperture and shows there is no significant contamination from other sources of IR emission and that the variability of X-1 is significantly greater than that of the background. The 3.6 and 4.5 $\mu$m fluxes of all available IRAC imaging of X-1 are provided (in $\mu$Jy) in Tab.~\ref{tab:Flux}, and the photometry is shown in Fig.~\ref{fig:LC}D. $1-\sigma$ flux uncertainties were determined from the background annulus and are consistent with the warm mission point-source sensitivity provided by the IRAC instrument handbook.



\section{Analysis and Discussion}

\subsection{Mid-IR Light Curve and Serendipitous, Near-Contemporaneous X-ray/Radio Observations}

The mid-IR light curve of X-1 (Fig.~\ref{fig:LC}D) reveals two events of significant mid-IR variability at both 3.6 and 4.5 $\mu$m in the observations taken on 2014 May 18 and 2015 Jun 4. At 3.6 $\mu$m, the 2014 May and 2015 Jun fluxes were $\sim60\%$ ($\sim4\sigma$) and $\sim90\%$ ($\sim6\sigma$) greater than the mean ``baseline" flux of the archival observations taken 2004 Oct 8 and 2004 Oct 10. The consistency of the 3.6 $\mu$m flux measured 4 months prior and 1 month after each variability event suggests both events exhibited a similar duration of $\sim5$ months. The 4.5 $\mu$m flux exhibits a similar trend as the 3.6 $\mu$m values except for the higher flux measured on 2015 Jul 2. Mid-IR fluxes between 2014 Jan and 2015 Jul are consistently $2\sigma$ higher than the baseline, whereas the three observations taken in Jan 2016 and the most recent one taken in Jan 2017 exhibit fluxes consistent with the baseline values.

Serendipitous, near-contemporaneous X-ray and radio observations of X-1 were performed by Cseh et al. (2015) within a $\sim$week of mid-IR observations of the first variability event taken on 2014 Jan 13 and 2014 May 19 (see Fig.~\ref{fig:LC}D). Although X-1 is known to exhibit variations in X-ray luminosity by up to a factor of 14 on timescales of 4 months (Gris{\'e} et al. 2010), \textit{Swift} observations taken between 2014 Jan 17-22  ($L_X=9.1\pm0.6\times10^{39}$ erg s$^{-1}$) and \textit{Chandra} observations taken on 2014 May 25 ($L_X=9.0\pm0.8\times10^{39}$ erg s$^{-1}$) did not reveal any significant variability in the 0.3-10 keV band (Cseh et al. 2015). These values are close to the 0.3-10 keV luminosity measured by \textit{XMM-Newton} on 2004 April 15 ($L_X=1.1\times10^{40}$ erg s$^{-1}$; Goad et al. 2006), which was obtained within 6 months prior to the baseline mid-IR observation. For the radio emission, it is difficult to infer any variability due to different bands used for each observation (1.6, 5, and 8-10 GHz) and the uncertainties in the spectral index (Cseh et al. 2014, 2015).


\subsection{Nature of the Mid-IR Emission}

Multiple explanations have been proposed for the mid-IR emission associated with X-1 and other ULXs (Berghea et al. 2010a; Dudik et al. 2016). Possible origins include heated dust shells like those detected around red supergiants, circumbinary dust disks, jets, and/or free-free emission from accretion disk winds. The latter mechanism was the proposed interpretation of the mid-IR emission from the luminous X-ray binary SS 433, a potential ULX-like source within the galaxy (Fuchs et al. 2006). In order to determine the nature of the mid-IR emission and flaring activity from X-1, we consider origins from a jet, disk outflow, and dust.

\begin{figure}[t!]
	\centerline{\includegraphics[scale=0.5]{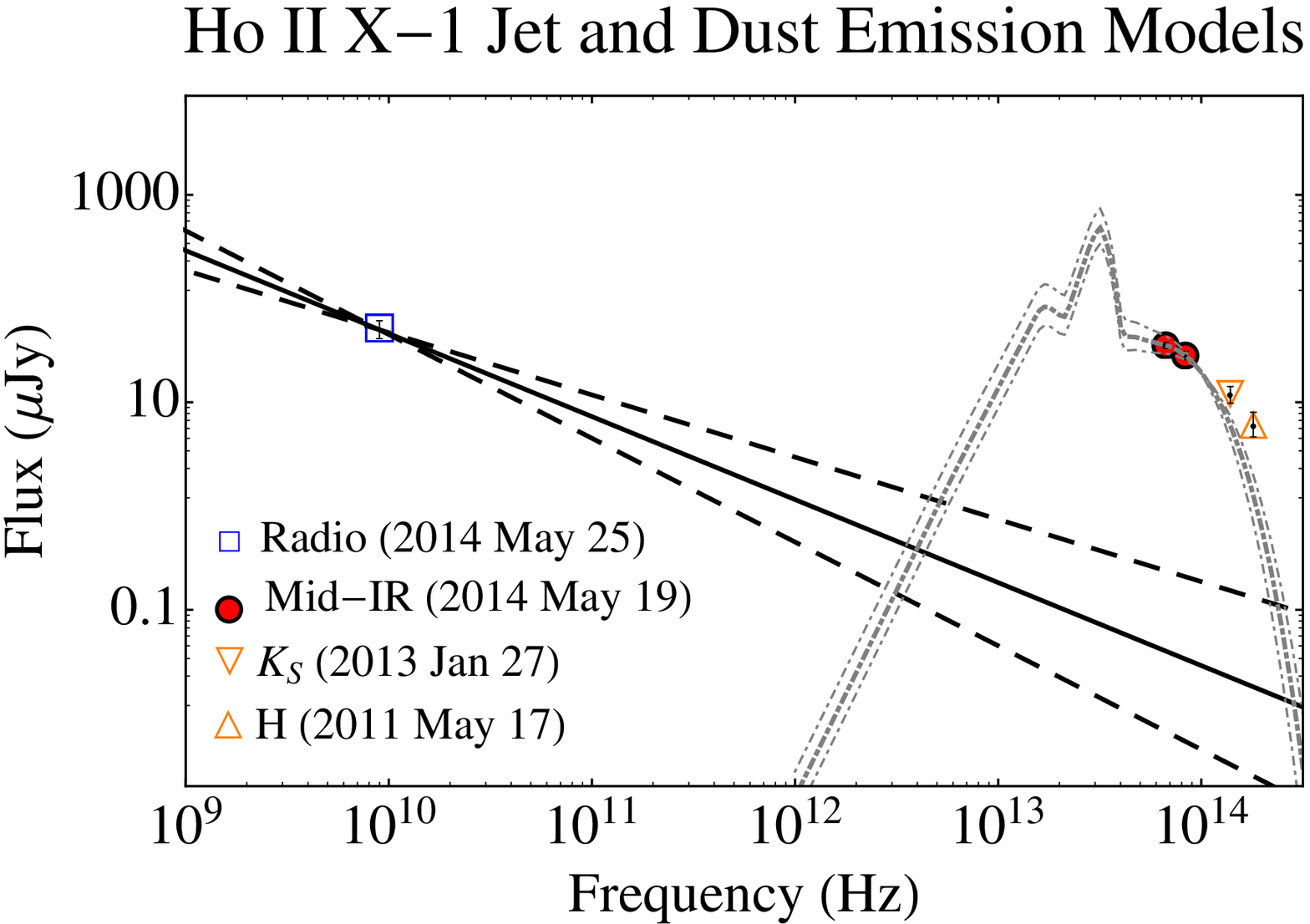}}
	\caption{Spectral energy distribution with near-contemporaneous X-band radio and mid-IR flux measurements of X-1. The $F_{\nu,\,\mathrm{jet}}\propto\nu^{\gamma}$ jet emission model from Cseh et al. (2014), where $\gamma=-0.8 \pm0.2$ (solid and dashed lines), is normalized to the X-band flux of the central X-1 radio source. The dot-dashed grey lines show the $a=0.1$ $\mu$m-sized silicate dust model fit to the 3.6 and 4.5 $\mu$m photometry with the $\pm1-\sigma$ uncertainty models shown in the thin dot-dashed grey lines. Orange triangles show the near-IR K$_S$- and H-band photometry of X-1 obtained by Heida et al. (2014).}
	\label{fig:H_SED}
\end{figure}

\subsubsection{Jet and Disk Outflow Hypothesis}

First we consider mid-IR emission from a jet, which is plausible since jet ejections have been inferred from Very Large Array (VLA) radio observations of X-1 (Cseh et al. 2014, 2015). The X-band (8 - 10 GHz) observations presented by Cseh et al. (2015) were taken a week after the mid-IR detection of the 2014 May peak. The near-contemporaneous nature of the mid-IR and radio measurement allows us to determine if the mid-IR emission can be reproduced by the power-law jet models fit to the radio flux. \footnote{The $\sim3 - 12$ yr adiabatic cooling timescales derived for the jet (Cseh et al. 2015) further validate this use of the near-contemporaneous mid-IR and radio data.} For the jet emission, we adopted the power-law fit by Cseh et al. (2014) with a spectral index of $\gamma=-0.8 \pm0.2$, where $F_{\nu,\,\mathrm{jet}}\propto\nu^{\gamma}$. This jet emission model normalized to the X-band flux of the central source ($F_X=50\pm10$ $\mu$Jy; Cseh et al. 2015) shows that the predicted mid-IR jet emission is over two orders of magnitude lower than the observed mid-IR flux (Fig.~\ref{fig:H_SED}). It is therefore unlikely that the mid-IR emission originates from the radio jet.


\begin{figure*}[t!]
	\centerline{\includegraphics[scale=1.3]{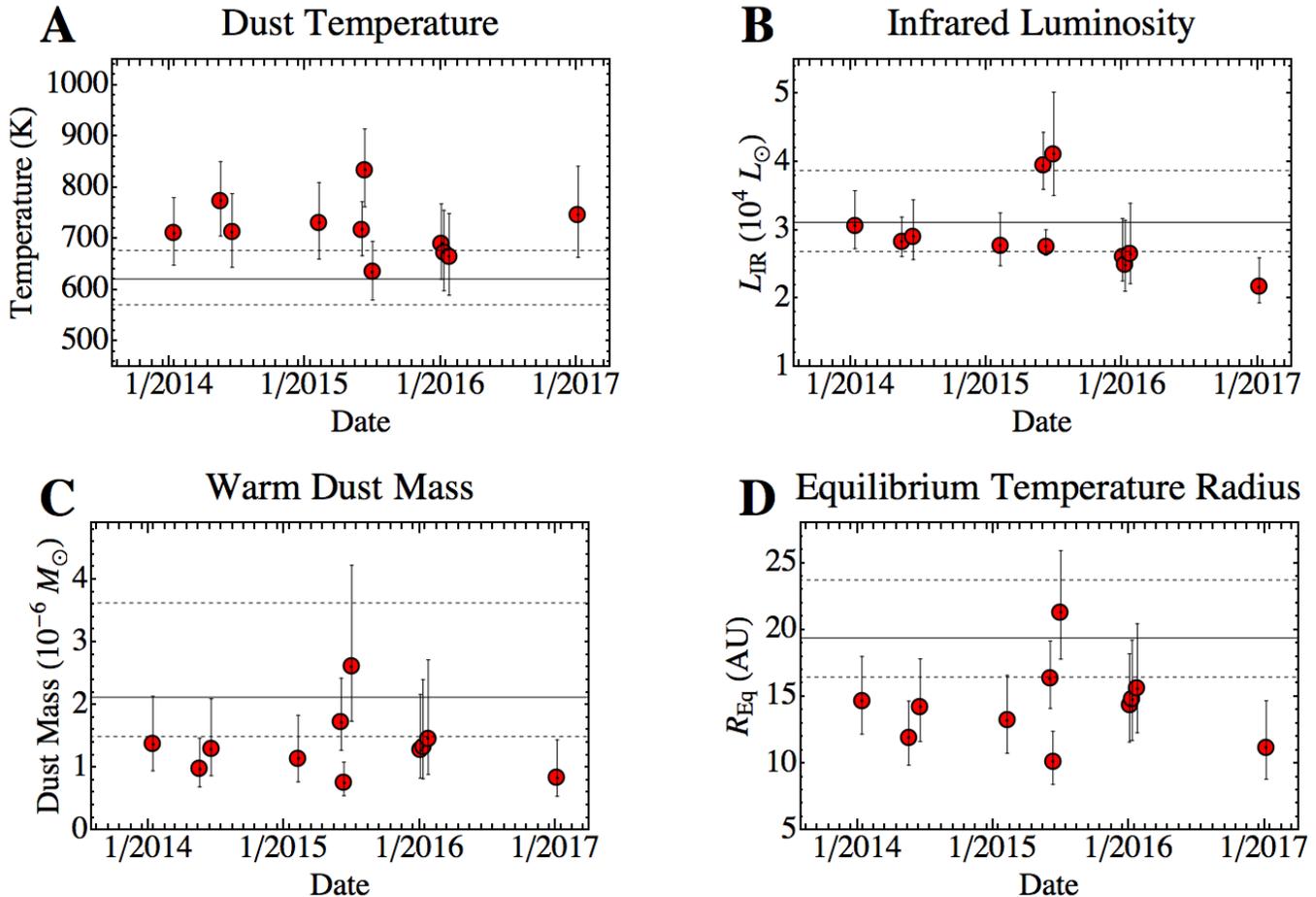}}
	\caption{Evolution of the (A) dust temperature, (B) IR luminosity, (C) dust mass, and (D) equilibrium temperature radius derived from the 3.6 and 4.5 $\mu$m fluxes. Solid and dashed lines indicate the mean baseline values and $1-\sigma$ uncertainty, respectively.}
	\label{fig:DustProp}
\end{figure*}

The disk outflow hypothesis is difficult to address due to the lack of mid-IR spectroscopy of X-1. Assuming the mid-IR flux from X-1 arises from free-free emission, it should exhibit a spectral index of $\alpha\sim-0.1$, where $F_{\nu,\,\mathrm{ff}}\propto\nu^{\alpha}$, if it is optically thick or $\alpha\sim0.6$ if it is optically thin. The decreasing mid-IR flux from X-1 as a function of frequency suggests it would be tracing optically thick emission; however, the mid-IR spectral index measured from the baseline and during the 2015 variability event are $\alpha=-2.26^{+0.85}_{-0.72}$ and $-1.38^{+0.37}_{-0.34}$, significantly steeper than the expected $\alpha\sim-0.1$ that is fit to the mid-IR emission from SS 433 (Fuchs et al. 2006). Additionally, at radio wavelengths X-1 is dominated by non-thermal emission (Cseh et al. 2014), which further disputes the presence of a significant free-free emission component.




\subsubsection{Warm Dust Hypothesis and Derived Dust Properties}


We suggest that the mid-IR counterpart of X-1 is dominated by dust continuum emission. A dust-hostile environment around X-1 might be inferred from the detection of high ionization lines such as [O IV] (Berghea et al. 2010a) with the \textit{Spitzer} Infrared Spectrograph (IRS; Houck et al. 2004). However, the large angular size of the aperture used by Berghea et al. (2010a; $\sim9''$, $\sim130$ pc) is dominated by the extended and bright nebulosity surrounding X-1. Additionally, the $\sim1$ mJy sensitivity of the high spectral resolution modes of \textit{Spitzer}/IRS were not likely able to detect the dust continuum emission from X-1. 

With high spatial resolution ($\sim0.7''$, $\sim10$ pc) and background-subtracted near-IR spectroscopy from Keck/MOSFIRE, Heida et al. (2016) detect a strong low-excitation [Fe II] ($\lambda=1.644$ $\mu$m) emission line that implicates the presence of a self-shielded region in the vicinity of X-1 where dust may also exist. Indeed, the mid-IR dust models described in this section fit the spectral shape of the mid- and near-IR photometry shown in Fig.~\ref{fig:H_SED} unlike the other emission models. We note that only the 3.6 and 4.5 $\mu$m fluxes were used to derive the fit since emission from sources in the ULX other than dust like the stellar photosphere of the donor star will start to contribute in the near-IR. Based on the red mid-IR color of X-1, photospheric emission is unlikely a significant contributor to the mid-IR flux.

Dust properties derived from the mid-IR photometry are summarized in Tab.~\ref{tab:Flux}. The $1-\sigma$ uncertainties of the dust properties were determined by applying the $\pm1-\sigma$ flux uncertainties to the parameter fits described in the following paragraphs.


The dust properties of the X-1 mid-IR counterpart over time can be determined from the measured 3.6 and 4.5 $\mu$m fluxes and provide insight on the mechanism that drives the emission. We note that longer wavelength coverage beyond 4.5 $\mu$m would provide more robust dust property calculations, however there are currently no platforms sensitive enough to detect such emission. The mid-IR dust emission is assumed to take the form $F_\lambda \propto Q_\mathrm{Sil}(\lambda,a )\,B_\lambda(T_d)$, where $F_\lambda$ is the flux at wavelength $\lambda$, $Q_\mathrm{Sil}(\lambda,a)$ is the grain emissivity model for silicate grains of radius $a$, and $B_\lambda(T_d)$ is the Planck function for a dust temperature, $T_\mathrm{d}$. Dust temperatures were derived by solving for $T_d$ from the 3.6 and 4.5 $\mu$m flux ratio

\beq
\frac{F_{3.6}}{F_{4.5}}=\frac{Q_\mathrm{Sil}(3.6\,\mu\mathrm{m},a )\,B_{3.6}(T_d)}{Q_\mathrm{Sil}(4.5\,\mu\mathrm{m},a )\,B_{4.5}(T_d)}
\eeq

\noindent assuming a single temperature component of optically thin $a=0.1$ $\mu$m-sized silicate grains. The dust temperatures range from $T_\mathrm{d}\sim600-800$ K (Fig.~\ref{fig:DustProp}A).


\begin{figure*}[t!]
	\centerline{\includegraphics[scale=1.3]{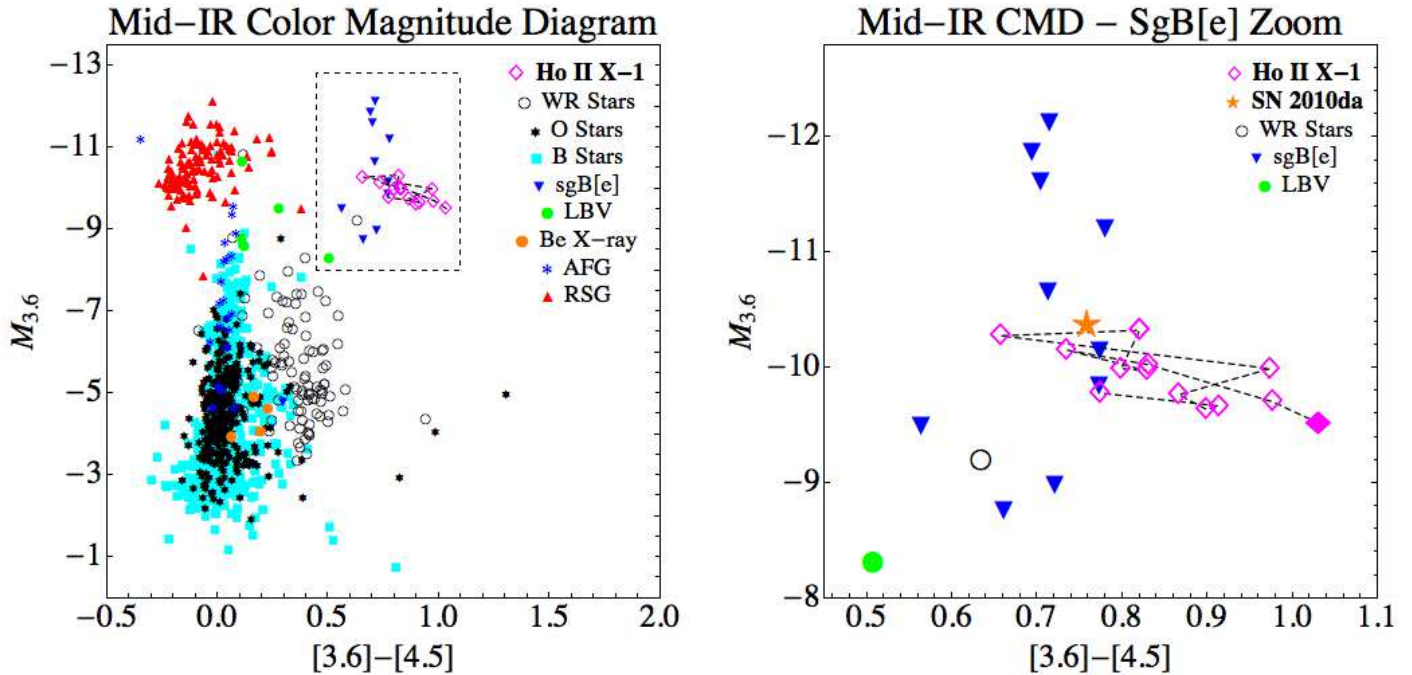}}
	\caption{ (Left) Mid-IR CMD of massive stars in the Large Magellanic Cloud (Bonanos et al. 2009) over plotted with measurements of X-1 (pink diamonds). (Right) Mid-IR CMD zoomed in on X-1, the population of sgB[e]-stars, and the sgB[e]-HMXB SN~2010da (orange star). The filled diamond corresponds to the baseline observation of X-1 taken on 2004 Oct 8.}
	\label{fig:CMD}
\end{figure*}

The IR luminosity, $L_\mathrm{IR}$, was estimated by integrating over the fitted dust emission model and assuming a distance of 3.39 Mpc to X-1 (Karachentsev et al. 2002). We find that $L_\mathrm{IR}$ does not exhibit substantial variations from the baseline value of $L_\mathrm{IR}=3.1^{+1.2}_{-0.7}\times10^4$ $\mathrm{L}_\odot$ (Fig.~\ref{fig:DustProp}B). 

The mass of the emitting dust, $M_\mathrm{d}$, was derived from the mid-IR flux and dust model as shown in Eq.~\ref{eq:Mass}:

\beq
M_d=\frac{(4/3)\,a\,\rho_b\,F_\lambda\,d^2}{Q_\mathrm{Sil}(\lambda,a )\,B_\lambda(T_d)},
\label{eq:Mass}
\eeq

\noindent
where $\rho_b$ is the bulk density of the dust grains and $d$ is the distance to the source. A bulk density of $\rho_b=3$ $\mathrm{gm}$ $\mathrm{cm}^{-3}$ (e.g. Draine \& Li 2007) is assumed for the emitting silicate grains. The dust mass exhibited minima around 2014 May and 2015 June with values of $M_\mathrm{d}=1.0^{+0.5}_{-0.3}\times10^{-6}$ $\mathrm{M}_\odot$ and $0.8^{+0.3}_{-0.2}\times10^{-6}$ $\mathrm{M}_\odot$, respectively (Fig.~\ref{fig:DustProp}C).

Assuming the dust is heated radiatively, the equilibrium temperature radius, $R_\mathrm{eq}$, can be derived as the distance from the heating source where the energy input from the radiation field is equal to the radiative output from the dust grains of temperature $T_\mathrm{d}$. The value of $R_\mathrm{eq}$ provides an estimate of the lower limit or inner radius of the surrounding warm dust and can be approximated by

\beq
R_\mathrm{eq}\approx\left(\frac{Q_\mathrm{abs}}{Q_\mathrm{e}}\frac{L_\mathrm{IR}}{16\pi\sigma T^4_d}\right)^{1/2},
\label{eq:temp}
\eeq
\noindent
where $Q_\mathrm{abs}$ and $Q_\mathrm{e}$ are the grain absorption and emission Planck mean cross sections (see Draine 2011), respectively, and $\sigma$ is the Stefan-Boltzmann constant. A typical value of $\left(Q_\mathrm{abs}/Q_\mathrm{e}\right)\approx0.3$ for a stellar heating source (e.g. Smith et al. 2016) is assumed for the calculation. Similar to the dust mass, $R_\mathrm{eq}$ exhibits local minima around 2014 May and 2015 June where $R_\mathrm{eq}=11.9^{+2.8}_{-2.0}$ AU and $10.1^{+2.3}_{-1.7}$ AU, respectively (Fig.~\ref{fig:DustProp}D). We suggest that dust is likely confined to a disk-like geometry given that the IR luminosity is $\sim4\%$ of X-1's average isotropic X-ray luminosity of $4\times10^{39}$ $\mathrm{L}_\odot$ (Kaaret et al. 2004). 

\subsection{X-1 Donor: A Supergiant B[e] Star?}

The mid-IR colors and absolute magnitude can be used to interpret the nature of the mid-IR counterpart of X-1 by comparing against a sample of various classes of identified massive stars. We utilize the Bonanos et al. (2009) sample of 1750 massive stars in the Large Magellanic Cloud (LMC) observed with \textit{Spitzer}/IRAC. The mid-IR color-magnitude diagram (CMD) shown in Fig.~\ref{fig:CMD} reveals that X-1 is consistent with the region of phase space occupied by supergiant (sg) B[e]-stars, which may be the stellar type of the mass donor in the system. The [3.6]-[4.5] colors of X-1 ($0.7-1.0$) are also consistent with the [3.6]-[4.5] colors exhibited by the population of sgB[e] stars in M31 and M33 due to the presence of warm circumstellar dust (Humphreys et al. 2016).


SgB[e] stars are B-type supergiants that exhibit forbidden emission lines and an IR excess attributed to circumstellar dust in a dense equatorial disk or torus (Lamers et al. 1998; Kastner et al. 2006; Clark et al. 2013). These stars are thought to be in a short lived post-main sequence phase in the evolutionary track of a massive star. The mechanism that produces the dense dust disks around sgB[e] stars is currently uncertain, but their enhanced equatorial mass-loss has been suggested to be linked with binary interaction (de Wit et al. 2014 and ref. therein). 


Previous spectroscopic observations of X-1 reveal properties that are support the sgB[e] interpretation. 
The presence of the [Fe II] ($\lambda=1.644$ $\mu$m) emission line, a series of hydrogen Brackett lines (Heida et al. 2016), and the mid-IR excess detected from X-1 are common characteristics of B[e] and sgB[e] stars (Lamers et al. 1994; Humphreys et al. 2016).  The resolved width of the [Fe II] line is $\sim120$ km $\mathrm{s}^{-1}$ (Heida et al. 2016), which is consistent with the $\sim100$ km $\mathrm{s}^{-1}$ equatorial outflows observed from sgB[e] stars (de Wit et al. 2014 and ref. therein). Lastly, strong [Fe II] and H-emission lines have also been detected from dust-obscured sgB[e] high-mass X-ray binary (HMXB) systems like IGR J16318-4848 (Filliatre \& Chaty 2004; Moon et al. 2007).


The derived dust properties (Fig.~\ref{fig:DustProp} and Tab.~\ref{tab:Flux}) are consistent with the sgB[e] hypothesis. The range of IR luminosities exhibited by X-1 ($L_\mathrm{IR}\sim3 - 4\,\times10^4$ $\mathrm{L}_\odot$) agree with the lower luminosity range of sgB[e] stars (several times $10^4$ $\mathrm{L}_\odot$). We note that our derived IR luminosity is a lower limit of the total luminosity of the heating source since we are only sensitive to warm ($\sim600$ K) dust at 3.6 and 4.5 $\mu$m.  Remarkably, the equilibrium temperature radius of surrounding dust ($R_\mathrm{eq}\sim10 - 20$ AU) matches the inner edge radii of spatially resolved galactic sgB[e] disks: $\lesssim30$ AU for CPD-52 9243 (Cidale et al. 2012) and $\sim 20 - 30$ AU for MWC 300 (Wang et al. 2012).

X-1 is therefore a member of an emerging class of luminous X-ray sources with donor stars that resemble sgB[e] stars such as IGR J16318-4848 (Filliatre \& Chaty 2004), GX 301-2 (Servillat et al. 2014), and SN~2010da (Binder et al. 2016; Lau et al. 2016; Villar et al. 2016). We note that a majority of known sgB[e] stars do not exhibit significant photometric or spectroscopic variability (e.g. Zickgraf et al. 1986); however, there are some sgB[e]s where significant photometric IR variability has been observed such as CI Cam (Thureau et al. 2009) and SN~2010da, both of which are HMXBs. This suggests that the mid-IR variability exhibited by these X-ray luminous systems may be linked to the presence of a close companion.


\subsection{Interpreting the Mid-IR Variability}

Mid-IR variability traced by dust emission at X-1 implicates fluctuations in the energetics and/or total mass of the emitting dust. The consistently higher dust temperature around 2014 May and 2015 June over the baseline (Fig.~\ref{fig:DustProp}A) and the lower dust masses (Fig.~\ref{fig:DustProp}C) suggest the mid-IR variability is due to increased dust heating as opposed to enhanced dust production. Due to the uncertainties of the derived dust properties, it is unclear whether the increased dust heating is caused by variations in the heating source luminosity and/or fluctuations in the geometry of the emitting dust. However, the consistent X-ray flux from X-1 reported by Cseh et al. (2015) around the first mid-IR variability event suggests it was not associated with X-ray activity from the ULX. 



Assuming the mass donor of X-1 is a sgB[e] star, we can gain further insight on the mid-IR variability and circumstellar dust geometry by estimating the binary separation, $a$, and orbital period, $P$. The luminous X-ray flux from X-1 requires significant mass transfer from the donor to the accretor that likely occurs through Roche Lobe overflow. The Roche lobe radius of the donor star, $R_L$, can be related to $a$ and the mass ratio of the donor and accretor, $q$, by the following expression (Eggleton 1983):

\beq
\frac{R_L}{a} = \frac{0.49\,q^{2/3}}{0.6\,q^{2/3}+\mathrm{Log}(1+q^{1/3})}.
\label{eq:a}
\eeq
\noindent
For the donor star, we adopt a mass and radius of M$_*=30$ M$_\odot$ and R$_*=27$ R$_\odot$, consistent with the sgB[e] star in IGR J16318-4848 (Filliatre \& Chaty 2004; Rahoui et al. 2008), and we assume that the accretor is a 25 M$_\odot$ black hole (Cseh et al. 2014). If $R_L=R_*$, we find that 

\beq
a\sim0.32\,\mathrm{AU}\, \,\mathrm{and}\,\,P\sim8.8\,\mathrm{d},
\label{eq:b}
\eeq
\noindent
where $P$ has been derived from Kepler's Third Law. Given that $a<R_\mathrm{eq}\sim10-20$ AU (Fig.~\ref{fig:DustProp}D), the dust is likely in a circumbinary disk/torus. Circumbinary dust confined to a torus may therefore be shielded from the ULX by the accretion disk surrounding the compact object, which could explain why the mid-IR variability does not appear to follow the X-ray luminosity. Due to the week- to month- timescales of the mid-IR observations, it is difficult to determine if the variability from the dust emission is linked to a $\sim9$-day orbital period. Ultimately, we leave the details of the physical processes causing increased dust heating as an open question and emphasize the importance of obtaining contemporaneous multi-wavelength observations as well as high-cadence ($\sim$ day timescale) IR observations.

\section{Conclusions and Outlook}

We presented \textit{Spitzer}/IRAC data from the SPIRITS survey revealing mid-IR variability from the ULX Holmberg II X-1, which we attribute to increased heating of dust in a circumbinary torus. We have also shown that the mid-IR counterpart and likely mass donor in X-1 resembles a sgB[e]-star based on its location in a mid-IR CMD, its near-IR spectral features, and the emitting dust properties. It is currently unclear if there is a link between the evolutionary state of sgB[e] stars and interactions with compact object companions, but the continued discovery of X-ray luminous sgB[e] systems with mid-IR counterparts holds exciting prospects for future multi-wavelength studies.  

In addition to continued X-ray/IR/radio monitoring, characterizing the emitting dust at wavelengths longer than 4.5 $\mu$m with mid-IR spectroscopy from upcoming IR platforms such as the James Webb Space Telescope will be crucial for exploring possible connections between circumbinary material and the ultra luminous X-ray activity from these enigmatic sources.



\emph{Acknowledgments}. 

R.L. would like to thank Kaew Tinyanont, Bob Gehrz, and the anonymous referee for the valuable comments and insight. R.L. would also like to thank Jacob Jencson, Michael Ressler, and the rest of the SPIRITS team. This work made use of observations from the \textit{Spitzer} \textit{Space Telescope} operated by the Jet Propulsion Laboratory, California Institute of Technology, under a contract with NASA (PIDS 10136, 11063, \& 13053). This work was partially carried out at the Jet Propulsion Laboratory, California Institute of Technology, under a contract with the National Aeronautics and Space Administration. 

\newpage

\clearpage

\begin{deluxetable}{cccccccc}
\tablecaption{Observed fluxes and dust properties of Holmberg II X-1}
\tablewidth{0pt}
\tablehead{ Date & MJD & $F_\mathrm{3.6}$  &$F_\mathrm{4.5}$ & $T_\mathrm{d}$ (K) & $\mathrm{M}_\mathrm{d}$ & $\mathrm{L}_\mathrm{IR}$  & $\mathrm{R}_\mathrm{eq}$ \\ (Year Month Day) && ($\mu$Jy) & ($\mu$Jy) & (K) & $(\times\,10^{-6}\,\mathrm{M}_\odot)$  & $(\times\,10^{4}\,\mathrm{L}_\odot)$& (AU) }

\startdata
\text{2004 October 08} & 53286.40 & 15.65 (2.71) & 25.89 (2.37) & $609_{-78}^{+86}$ & $2.2_{-1.}^{+2.6}$ & $3.1_{-0.7}^{+1.2}$ & $20_{-5}^{+7}$ \\
 \text{2004 October 10} & 53288.31 & 18.62 (2.45) & 29.28 (2.45) & $634_{-65}^{+71}$ & $2._{-0.8}^{+1.6}$ & $3.2_{-0.5}^{+0.9}$ & $19_{-4}^{+5}$ \\
 \text{2014 January 13} & 56670.31 & 25.05 (2.54) & 34.44 (2.37) & $711_{-63}^{+69}$ & $1.4_{-0.4}^{+0.8}$ & $3.1_{-0.3}^{+0.5}$ & $15_{-2}^{+3}$ \\
 \text{2014 May 19} & 56796.30 & 28.18 (2.62) & 35.45 (2.28) & $774_{-69}^{+76}$ & $1._{-0.3}^{+0.5} $& $2.8_{-0.2}^{+0.4}$ & $12_{-2}^{+3}$ \\
 \text{2014 June 19} & 56827.29 & 23.86 (2.62) & 32.75 (2.71) & $712_{-68}^{+76}$ & $1.3_{-0.4}^{+0.8}$ & $2.9_{-0.3}^{+0.5}$ & $14_{-3}^{+4}$ \\
 \text{2015 February 08} & 57062.00 & 24.28 (2.62) & 32.41 (2.37) & $731_{-71}^{+79}$ & $1.1_{-0.4}^{+0.7}$ & $2.8_{-0.3}^{+0.5}$ & $13_{-2}^{+3}$ \\
 \text{2015 June 04} & 57177.29 & 33. (2.62) & 44.93 (2.37) & $717_{-51}^{+55}$ & $1.7_{-0.4}^{+0.7}$ & $3.9_{-0.3}^{+0.5}$ & $16_{-2}^{+3}$ \\
 \text{2015 June 11} & 57184.31 & 31.56 (2.62) & 36.98 (2.45) & $834_{-72}^{+80}$ & $0.8_{-0.2}^{+0.3}$ & $2.8_{-0.1}^{+0.2}$ & $10_{-2}^{+2}$ \\
 \text{2015 July 02} & 57205.45 & 24.28 (2.71) & 38.08 (2.37) & $635_{-55}^{+60}$ & $2.6_{-0.9}^{+1.6}$ & $4.1_{-0.6}^{+0.9}$ & $21_{-3}^{+5}$ \\
 \text{2016 January 03} & 57390.05 & 19.72 (2.37) & 28.01 (2.28) & $690_{-70}^{+77}$ & $1.3_{-0.4}^{+0.9}$ & $2.6_{-0.3}^{+0.6}$ & $14_{-3}^{+4}$ \\
 \text{2016 January 10} & 57397.52 & 17.52 (2.37) & 25.64 (2.28) & $673_{-74}^{+83}$ & $1.3_{-0.5}^{+1.1}$ & $2.5_{-0.4}^{+0.7}$ & $15_{-3}^{+4} $\\
 \text{2016 January 24} & 57411.48 & 18.02 (2.54) & 26.74 (2.54) & $665_{-76}^{+84}$ & $1.5_{-0.6}^{+1.3}$ & $2.6_{-0.4}^{+0.7}$ & $16_{-3}^{+5}$ \\
 \text{2017 January 05} & 57758.51 & 19.97 (2.45) & 26.06 (2.12) & $747_{-83}^{+95}$ & $0.8_{-0.3}^{+0.6}$ & $2.2_{-0.2}^{+0.4} $& $11_{-2}^{+3}$ \\
\enddata

\tablecomments{Fluxes at 3.6 and 4.5 $\mu$m are indicated by $F_\mathrm{3.6}$ and $F_\mathrm{4.5}$. $T_\mathrm{d}$, $\mathrm{M}_\mathrm{d}$,  $\mathrm{L}_\mathrm{IR}$, and $\mathrm{R}_\mathrm{eq}$ are the dust temperature, dust mass, IR luminosity, and equilibrium temperature radius, respectively. $1-\sigma$ flux uncertainties are provided in the parentheses.}
	\label{tab:Flux}
\end{deluxetable}

\end{document}